# Introduction of Sr into $Bi_2Se_3$ thin films by molecular beam epitaxy

L. Riney,[1] C. Bunker,[1] S.-K. Bac,[1] J. Wang,[1] D. Battaglia,[1] Yun Chang Park[2], M. Dobrowolska[1], J.K. Furdyna[1], X. Liu,[1] B.A. Assaf[1]

[1]Department of Physics, University of Notre Dame, Notre Dame IN, 46556, USA
[2]Department of Division of Measurement & Analysis, National Nanofab Center, Daejeon 34141, Republic of Korea.

**Abstract.** $Sr_xBi_2Se_3$ is a candidate topological superconductor but its superconductivity requires the intercalation of Sr by into the van-der-Waals gaps of $Bi_2Se_3$. We report the synthesis of $Sr_xBi_2Se_3$ thin films by molecular beam epitaxy, and we characterize their structural, vibrational and electrical properties. X-ray diffraction and Raman spectroscopy show evidence of substitutional Sr alloying into the structure, while transport measurements allow us to correlate the increasing Sr content with an increased n-type doping, but do not reveal superconductivity down to 1.5K. Our results suggest that Sr predominantly occupies sites within a quintuple layer, simultaneously substituting for Bi and as an interstitial. Our results motivate future density functional studies to further investigate the energetics of Sr substitution into $Bi_2Se_3$.

The Bi-chalcogenides are the most prominent topological materials studied to date. Alloying transition and alkali metals into the Bi-chalcogenides has enabled the realization of the quantum anomalous Hall effect [1] and the discovery of topological superconductor candidates Sr-, Nb-, and Cu- doped $Bi_2Se_3$. [2] [3] [4] The latter has gained particular attention in the recent literature. $Cu-Bi_2Se_3$ was identified as a superconductor as early as 2010. [3] This was followed by the discovery of a rotational symmetry breaking in the superconducting state of Sr-, Nb, and Cu- doped $Bi_2Se_3$ evidenced by Knight shift measurements and by direct measurements of an anisotropic critical field. [4] [5] [6] [7] [8] This type of superconductivity is referred to as nematic. The nematicity identified in these materials is suggested as evidence of the topological superconductor phase predicted by Berg and Fu in $Bi_2Se_3$. [9], [10] Despite the encouraging results obtained from these measurements, tunneling experiments did not observe any evidence of topological superconductivity in $CuBi_2Se_3$ which has led to an ongoing controversy. [11]

All of the above work was done on bulk single crystals of Sr-, Nb, or Cu- doped $Bi_2Se_3$ as thin films of these materials are highly challenging to grow. Synthesis of such films can enable a wide variety of devices, including junctions and interferometers that can directly probe and manipulate Majorana modes. [12] [13] But, their properties sensitively depend on the location of Sr or Cu atoms in the lattice. Sr and Cu atoms can substitutionally replace Bi or Se, they can be intercalated within the van-der-Waals gaps of the structure or can be interstitial within a quintuple layer (see Fig. 1). Importantly, according to previous work, to become superconducting, crystals require intercalation within the Van-der-Waals gap according to previous work. [14] [3] However, Early work on $Cu-Bi_2Se_3$ grown by molecular beam epitaxy did not report any superconductivity. [15]

In this work, we synthesize $Sr_xBi_2Se_3$ thin films by molecular beam epitaxy on GaAs(111)B substrates. While we did not succeed in obtaining superconductivity, our results provide insight on the mechanism by which Sr is alloyed into $Bi_2Se_3$. First, X-ray diffraction measurements indicate an increasing c-axis lattice parameter, inconsistent with what is observed in single crystals with intercalated Sr. [14] Second, phonon linewidths broadening extracted from Raman spectroscopy indicates that Sr likely alloys into the structure. Third, an increasing n-type doping is seen with increasing Sr content, in addition to a marked decrease in mobility, indicating that Sr introduces donors. This leads us to hypothesize that Sr likely occupies multiple lattice sites, possibly substitutional and interstitial, where it can act overall as a donor. Our results indicate that the thermodynamics of MBE co-deposition at low temperature likely favor substitutional and interstitial alloying, thus making the synthesis of superconducting films with intercalated Sr highly challenging.

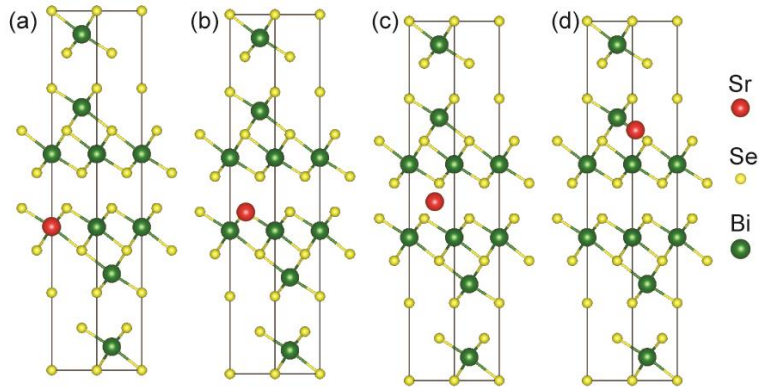

FIG 1. Lattice sites where Sr can be introduced into $Bi_2Se_3$ structure. (a) Bi substitution, (b) Se substitution, (c) Intercalation into van der Waals gap, (d) interstitial with a quintuple layer.

Synthesis

$Sr_xBi_2Se_3$ thin films are synthesized by molecular beam epitaxy (MBE) on GaAs(111)B substrate. Two sample series are studied, one grown at low temperature (200-240°C) and another at high temperature (280-300°C). The sample thickness is fixed at 70nm. The relevant growth parameters including the Sr cell temperature and the substrate temperature are listed in table I.

Results

The Sr content of each sample is determined by performing energy dispersive X-ray measurements. The content ranges from 0% to 22% across the 11 samples. The relative concentration of each element is calculated using the convention outlined in [16]. X-ray diffraction measurements are performed to extract the c-lattice constant of the layers for different values of x. A zoom-in on the (006) and the (0015) Bragg peaks of $Bi_2Se_3$ is shown in Fig. 2(a) and the lattice constant is plotted versus "x" in Fig. 2(b). The lattice constant of pristine $Bi_2Se_3$ agrees with what is typically measured in MBE samples (c=28.56Å). A clear monotonic increase in c is obtained as more Sr is alloyed into the structure. In bulk single crystals, the c-axis lattice constant also increases when x approaches 10%, however, the slope of that increase is smaller than what we observe. [17] [14] This agrees with previous work on $Sr_xBi_2Se_3$ [17]. In single crystals, Sr is intercalated in the van-der-Waals gaps. Our observations indicate that in MBE grown samples, Sr is most likely not intercalated. Sr is expected to replace Bi in the $Bi_2Se_3$ structure, however, it is surprising that this substitution can yield such a dramatic change in c, given that atomic radius of Sr is almost equal to that of Bi. We also note that a phase separation is observed at about x=0.22, thus setting a previously unknown solubility limit for Sr alloying into $Bi_2Se_3$ by MBE.

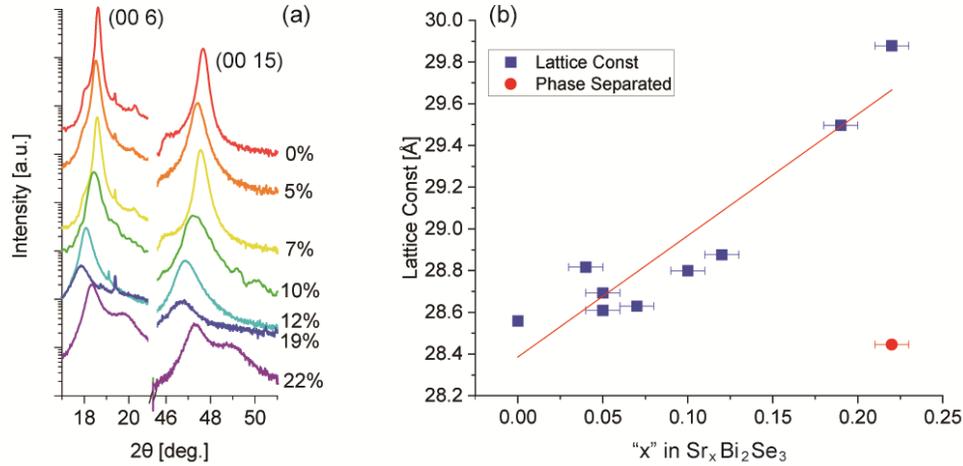

**FIG 2.** (a) Shift of the (006) and (0015) Bragg peaks with increasing Sr concentration. (b) Lattice constant versus Sr concentration '"x".

| "x" from EDS | T Sr | T subs | Lattice constant |
|---|---|---|---|
| 0 | 0 | 240 | 28.56 |
| 0.05 | 303 | 240 | 28.69 |
| 0.12 | 330 | 240 | 28.88 |
| 0.22 | 350 | 200 | 29.88* |
| 0.05 | 303 | 300 | 28.61 |
| 0.06 | 303 | 280 | 28.66 |
| 0.07 | 303 | 280 | 28.63 |
| 0.10 | 320 | 280 | 28.80 |
| 0.19 | 350 | 300 | 29.50 |
| 0.22 | 340 | 280 | 29.20* |

**Table 1.** Composition, MBE parameters and c lattice constant for the samples studied in this work. (*) indicates samples where phase separation was observed. x in $Sr_xBi_2Se_3$ is determined by energy dispersive X-ray spectroscopy (EDS).

Transmission electron microscopy images are taken on a set of samples selected to represent each composition range. Measurements from two samples with x=0.1 (Fig. 3(a,b)) and x=0.22 (Fig. 3(c,d)) are shown. In Fig. 3(a,b), the x=0.1 sample exhibits a uniform structure with quintuple layers clearly visible. Thus, despite the high concentration of Sr, $Bi_2Se_3$ is able to retain its crystal structure. In Fig. 3(c,d), TEM images taken on the x=0.2 sample are shown. A high magnification image in Fig. 3(c) shows a region of the sample where the quintuple layer structure is lost, likely transforming into a $Bi_xSe_y$-type structure, at high Sr content. Note that SrSe crystallizes in a rocksalt structure, [18] however, this structure is not observed in Fig. 3(c). Rather, the crystal resembles the R-3m space group, but with a loss of the 5-fold layering (2 Bi layers) seen in Fig. 3(a). We thus conclude, that this structure is likely a $(Sr,Bi)_xSe_y$ type segregate with a different stacking than $Bi_2Se_3$ but a comparable lattice parameter, similar to has been reported for Mn, Pb and Sn in $Bi_2Te_3$. [19] A number of these structures can form in the case of the selenide. [20] Fig. 3(d) shows another region of that same sample where the QL structure is retained, confirming the observed phase splitting seen in X-ray diffraction.

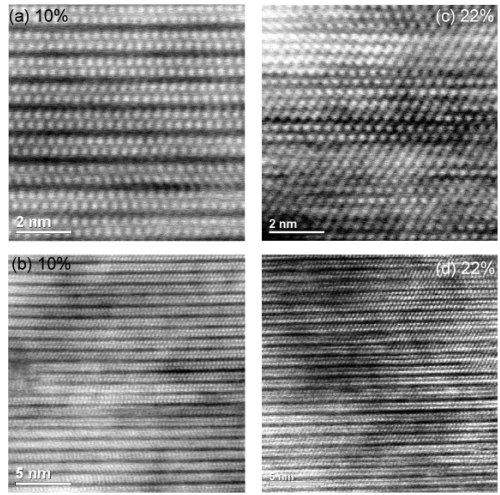

**FIG 3.** TEM images of two samples. (a) A high-resolution image of a small region on a 10% sample shows nearly pristine quintuple layering. (b) A larger region of the same sample showing minimal variation to the layered structure, evidencing a lack of intercalation. (c) A high-resolution image of a small region on a 22% sample, where the layering structure is much less prominent. (d) A larger region of the same sample showing the layered structure but with much more variation, evidencing some type Sr incorporation into the structure.

To get more insight into the mechanism by which Sr is introduced into the structure, we perform, Raman spectroscopy measurements performed using a green laser at room temperature. Fig. 4(a) shows 6 characteristic spectra taken from representative samples that are all single phase. The three expected $Bi_2Se_3$ phonon peaks are observed. [21] Qualitatively, the $E_g$ peak exhibits a slight redshift. All peaks are also broadened as Sr content is increased. The changing Lorentzian peak width versus x is shown in Fig. 4(b). The broadening is monotonic and breaks down upon the occurrence of the phase splitting at x=0.22. The Raman measurements further confirm our hypothesis that Sr is incorporated substitutionally into the structure. This result comes in contrast with recent Raman measurements on Cu-intercalated $Bi_2Te_2Se$ crystals where a robust preservation of the linewidth was observed, even up to high Cu concentrations. [22] Qualitatively, the broadening that we observe is similar to what has been reported in other alloys including (Al,Ga)As [23] and transition metal dichalcogenides [24] [25] where substitutional disorder is known to occur.

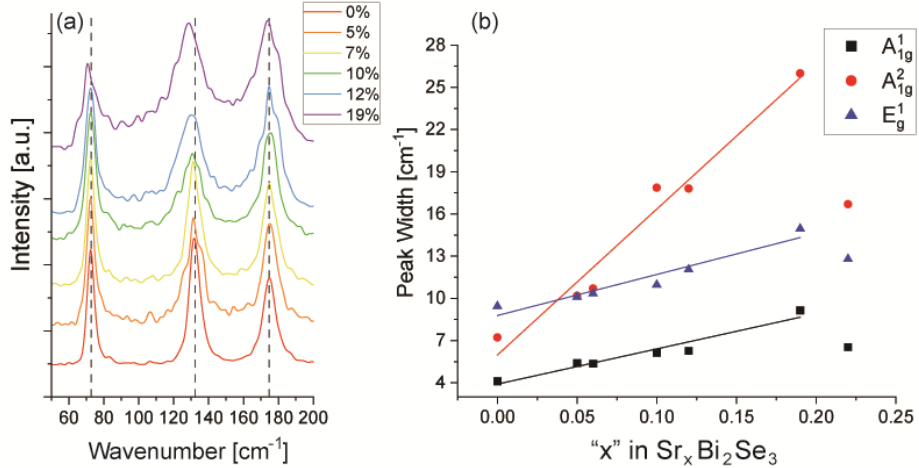

**FIG 4**. (a) Raman peak broadening versus Sr concentration suggesting alloying is increasing with x. (b) Spectra across sample composition show relatively constant peak position. The dotted lines indicate the three Raman-active phonon energies expected for $Bi_2Se_3$. Measurements were performed using a green 532nm laser.

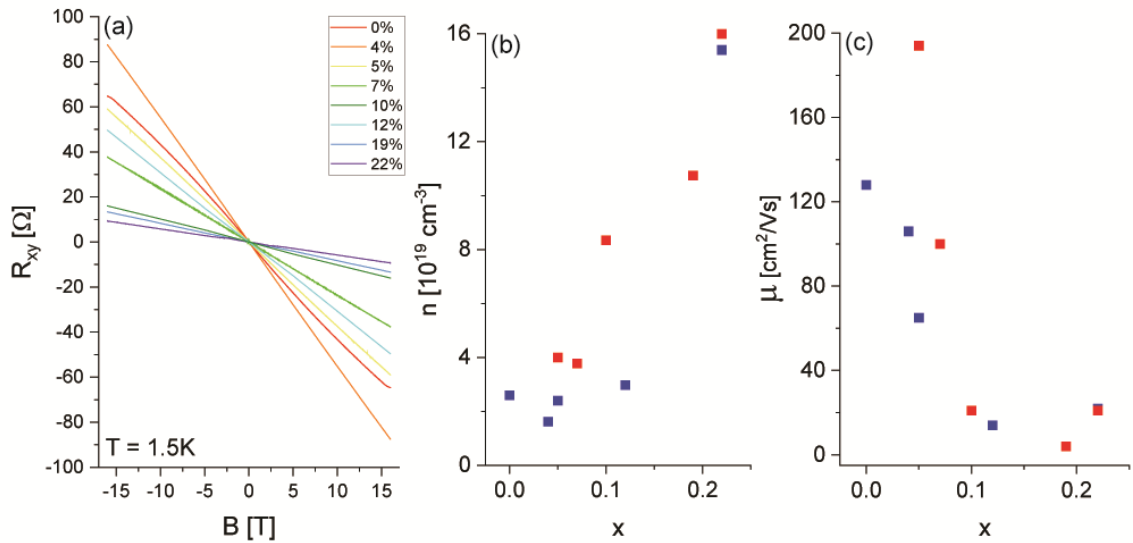

**FIG 5.** (a) Hall resistance versus magnetic field at 1.5K. (b) Carrier density and (c) mobility versus "x" extracted from the slope in of $R_H$. The blue and red squares represent samples grown at low and high substrate temperature respectively.

Strong evidence from our Raman spectroscopy and XRD measurements indicates that Sr atoms alloy substitutionally into $Bi_2Se_3$ during co-deposition by MBE. The most likely site for Sr to occupy is the Bi site, given the comparable valence of the two atoms ($Bi^{3+}$ and $Sr^{2+}$) yielding an acceptor type defect level. However, our electrical transport measurements indicate otherwise. These measurements are performed on rectangular cleaved pieces at 1.5K in magnetic fields up to 16T to extract the carrier density and mobility up to 16T and down to 1.5K. The Hall resistance extracted from all the studied samples is shown in Fig. 5(a). All curves are linear indicating that a single carrier type dominates. The carrier density is plotted in Fig. 5(b). Two colors are used to distinguish samples grown at different substrate temperatures. Low temperature growths yield a lower carrier density overall. However, the carrier density n systematically increases with increasing Sr content. At low "x", n~$10^{19}$cm$^{-3}$ is typical for MBE grown $Bi_2Se_3$. [26] In our $Sr_xBi_2Se_3$ thin films, it exceeds $10^{20}$cm$^{-3}$ before the crystal undergoes a phase separation. The mobility plotted in Fig. 5(c) drops dramatically as n increases, an expected behavior in the presence of charged impurities.

Previous work on bulk single crystals suggested that cations such as Sr, Ca and Cu can either substitute Bi at low concentrations or can intercalate at high concentrations. [27] [28] Our measurements of the increased n-type doping with increasing Sr rules out the first option, as Sr replacing Bi should generate p-type doping. The absence of superconductivity leads us to also hypothesize that Sr is mainly not intercalated during MBE co-deposition. It is possible that Sr is introduced at other lattice sites possibly interstitially into a quintuple, as suggested for Cu in a previous work. [29]

A photoemission study corroborates our result that Sr introduces electrons into the system. [30] This study hypothesized that Sr induces superconductivity and introduces donors both by substituting Bi and intercalating into the van-der-Waals gap. It suggested searching for intercalated Sr by TEM. Within our current resolution, we did not observe any intercalated species. While we cannot completely rule out intercalation, we can conclusively say that Sr can occupy other sites in the unit cell, through the combined observation of increased n-doping and increasing lattice constant with increasing x. Our results in the context of previous results on $Sr_xBi_2Se_3$ call for a more in-depth investigation of the formation energy of different types of Sr defects in the unit cell. We note that even when the Sr dopant density exceed $10^{21}$cm$^{-3}$ (10%), our carrier density does not exceed $10^{20}$cm$^{-3}$ electrons. Thus, each Sr contributes less than 0.1electrons, indicating that the atom can occupy different lattice sites that allow it to be a donor, an acceptor or a neutral defect. Density functional theory calculations on transition metal defects in $Bi_2Se_3$ in fact show that in addition to substituting a Bi atoms, Fe, Mn or Cr can also occupy interstitial sites within the quintuple layers. [31] These two lattice sites are found to have the lowest activation energies, making them the most energy favorable. Our results indicate that the thermodynamics of Sr defects in $Bi_2Se_3$ can be similarly explained.

Conclusion

In conclusion, our measurements combining structural, vibrational and electrical characterization lead us to conclude that Sr in MBE-grown $Sr_xBi_2Se_3$ occupies multiple sites in the unit cell, mostly likely the Bi site and an interstitial site. Intercalated Sr between the van-der-Waals layers is likely not the most favorable one. The strongest evidence of this comes from the observation that MBE samples grown by co-deposition exhibit a large change in the lattice constant that is not observed in intercalated crystals, [16] and do not host superconductivity down to 1.5K. The strong n-type doping observed in transport measurements rules out Sr occupation of Bi sites as dominant, leading us to believe that the thermodynamics of Sr entering interstitially need to be considered. Despite the absence of a superconducting transition down to 1.5K, our results suggest that MBE co-deposition enables the formation of a Sr$_x$Bi$_2$Se$_3$, possibly relevant for thermoelectric applications. We further showed that this alloy is stable and single phase up to x=0.22, thus establishing a previously unknown solubility limit for Sr in Bi$_2$Se$_3$.

**Sample preparation**

Sr$_x$Bi$_2$Se$_3$ thin films are synthesized by molecular beam epitaxy (MBE) on GaAs(111)B substrate. The native oxide is initially desorbed from the substrate surfaces. The growth is performed while maintaining a high Se:Bi ratio with the Bi cell maintained at 500°C and the Se cell at 180°C. Sr is co-evaporated during the growth. Varying the Sr cell temperature between 0 and 350°C allows us to tune the amount of Sr alloyed into the structure. Two sample series grown at low (240°C and 200°C) and high temperature (280°C to 300°C) are studied.

**Experimental methods**

The sample composition is determined using energy dispersive X-ray spectroscopy in a scanning electron microscope. X-ray diffraction is performed using a standard 6-angle high resolution diffractometer equipped with a monochromated Cu-K$\alpha$ source. Transmission electron microscopy (TEM) images are taken at room temperature both in TEM and scanning-TEM mode. Raman spectroscopy measurements are carried out at room temperature using a microscopic Raman spectrometer equipped with a green 523nm excitation laser. Lastly, magnetotransport measurements are performed at 1.5K and up to 16T using a liquid helium cryostat (Oxford Instruments) at an excitation current of $50\mu A$. The samples that are measured are rectangularly shaped. They are connected with 5 electrical wires using silver paste to simultaneously carry out Hall effect and magnetoresistance measurements.

Acknowledgements. Work supported by NSF-DMR-1905277.